\documentclass[12pt]{article}

\usepackage{amssymb}

\begin{document}

\begin{titlepage}

\begin{flushright}
\end{flushright}
\vskip 2.5cm

\begin{center}
{\Large \bf Absence of Long-Wavelength Cerenkov Radiation \\
With Isotropic Lorentz and CPT Violation}
\end{center}

\vspace{1ex}

\begin{center}
{\large Brett Altschul\footnote{{\tt baltschu@physics.sc.edu}}}

\vspace{5mm}
{\sl Department of Physics and Astronomy} \\
{\sl University of South Carolina} \\
{\sl Columbia, SC 29208} \\
\end{center}

\vspace{2.5ex}

\medskip

\centerline {\bf Abstract}

\bigskip

Modified theories of electrodynamics that include violations of
Lorentz symmetry often allow for the possibility of vacuum Cerenkov
radiation. This phenomenon has previously been studied in a number of
Lorentz-violating theories, but none of the methods that have previously
been developed are sufficient to study a theory with a timelike
Chern-Simons term $k_{AF}$, because such a term may generate exponentially growing
solutions to the field equations. Searching for vacuum Cerenkov radiation
in a theory with a purely timelike Chern-Simons term using only elementary
methods, we find that, despite the presence of the runaway modes, a charge in
uniform nonrelativistic motion does not radiate energy, up to second order in
the velocity.

\bigskip

\end{titlepage}

\newpage

In recent years, the closely related possibilities of Lorentz and CPT
symmetry violations have gotten an increasing amount of attention. Theories
that lack one or both of these symmetries, which are ordinarily considered
fundamental building blocks of the laws of physics, might describe new
physics related to quantum gravity. In fact, many theories that have been
proposed as candidates to explain quantum gravity suggest the possibility
of Lorentz symmetry breaking, at least in certain regimes. Moreover,
if any violation of these
symmetries were uncovered experimentally, it would be a discovery of
critical importance and would open up a new window for studying
fundmental physics.

Even if Lorentz and CPT violation are not realized in
nature (and, as yet, there is no compelling evidence that they are), studying
the kinds of theories that contain such symmetry violations may provide
important new insights into the behavior of quantum fields. There is an
effective quantum field theory, known as the standard model extension (SME), 
which has been developed to accomodate Lorentz and CPT violation in a general
fashion. The SME action contains operators that may be constructed from
standard model fields~\cite{ref-kost1,ref-kost2}.
These operators are much more general than those that
appear in the action for the standard model itself, because they are not
constrained by Lorentz invariance. However, some other restrictions on
the operators are still necessary in order to make the theory tractable.
Such restrictions typically include locality, superficial renormalizability,
and gauge invariance.

In this paper, we shall be interested in the quantum electrodynamics sector
of the SME. Although there may also be Lorentz-violating modifications to
the behavior of matter
fields, we shall be chiefly interested in a form of Lorentz and
CPT violation that affects the purely electromagnetic sector.
This is one of the most interesting terms that appears in the SME---the
electromagnetic Chern-Simons term. This term defines a
preferred spacetime direction, which may be spacelike or timelike.
The Chern-Simons term screens static fields over long distances and
splits the dispersion relations for right-handed and left-handed
electromagnetic waves. This birefringence effect has been searched for
and not seen. Polarized light that has traversed even cosmological
distances shows no sign of the kind of systematic rotation of
polarization that would result from the presence of the Chern-Simons
term~\cite{ref-carroll1,ref-carroll2,ref-kost11,ref-kost21},
and this produces numerical constraints that
are extremely strong. Yet despite the tightness of the empirical bounds,
the term is still extremely interesting, and understanding the behavior of
quantum electrodynamics with a Chern-Simons term can reveal new insights
about the general structure of field theory. For example,
the Chern-Simons term in the Lagrange density is not actually gauge invariant,
although the integrated action associated with the term is. Because of
this subtlety, the radiative corrections to the Chern-Simons terms are rather
complicated, and for a while they were quite controversial~\cite{ref-coleman,
ref-jackiw1,ref-victoria1,ref-chung1,ref-chung2,ref-chen,ref-chung3,ref-victoria2,
ref-andrianov,ref-altschul1,ref-ebert,ref-altschul2}.

Another subject that is of very general interest in theories with Lorentz
violation is the possibility of vacuum Cerenkov radiation. Ordinary
Cerenkov radiation is emitted by charged particles moving in matter, where
the equations of electrodynamics differ from their vacuum forms and Lorentz
symmetry is lost. In a Lorentz-invariant vacuum, a Cerenkov process such as
$e^{-}\rightarrow e^{-}+\gamma$ is forbidden by energy-momentum conservation.
However, processes that are kinematically forbidden when Lorentz symmetry is
exact may become allowed when this symmetry is weakly broken.
In theories in which the dispersion relations for particles differ from
their conventional relativistic forms, the
Cerenkov process could occur. So for any
Lorentz-violating modification of electrodynamics, two very natural questions to
ask are, {\em Under what circumstances does the theory permit vacuum Cerenkov
radiation?} and, {\em What are the properties of the radiation emitted in the
Cerenkov process?}

The possibility of vacuum Cerenkov radiation in SME electrodynamics
has been studied in significant detail, but questions still remain.
This paper will address the most important outstanding problem
in this area, which has actually represented a glaring deficiency of our
understanding of Lorentz-violating Cerenkov physics.
Previous analyses of vacuum Cerenkov processes
have used a variety of techniques. Some of the methods involved the use of
Green's functions---either directly, in combination with a
prescribed current source~\cite{ref-lehnert2},
or through the use of Feynman diagrams~\cite{ref-kaufhold}.
Other methods used a transformation to carry the Lorentz violation
into the charged matter sector; then radiation occurs when the
matter-sector Lorentz violation leads to faster-than-light motion.
Depending on the theory involved, it may be possible to use this
method globally~\cite{ref-altschul9} or only locally in Fourier
space (by looking at whether the motion of a charged particle
exceeds the phase speed of light for a particular propagation
mode~\cite{ref-altschul12,ref-anselmi}).

None of these methods, however, is really equipped to deal with
what happens at very long wavelengths when the electromagnetic
Lorentz violation is isotropic and odd under CPT. The methods discussed
above have different levels of usefulness, depending on the nature
of the Lorentz violation involved. However, they all
ultimately rely on an understanding of the modified dispersion
relations for electromagnetic excitations. In fact, the radiation rates
in all the situations previously studied
can be estimated fairly accurately simply by looking at the phase space
available for the process, using the correctly modified dispersion
relations.

The use of methods based on understanding the electromagnetic
dispersion relations leaves a major
gap in our understanding of these theories, because in certain
cases, the Lorentz violation may lead to a fundamental change in the
electromagnetic propagation structure at long wavelengths. Rather
than traveling waves with real frequencies, there may be
long-wavelength runaway modes that grow exponentially with time.
The instability itself may be cured by using an acausal Green's
function, but this obviously introduces other complications, and
it may not be clear how exactly the physical theory should be
defined. However, we shall see that it is still possible to learn
quite a bit about the possibility of vacuum Cerenkov radiation in this
regime, without being troubled by such potential ambiguities.

The electromagnetic Lagrange density, including all power-counting
renormalizable Lorentz-violating terms that can appear solely in the
photon sector is
\begin{equation}
{\cal L}=-\frac{1}{4}F^{\mu\nu}F_{\mu\nu}
-\frac{1}{4}k_{F}^{\mu\nu\rho\sigma}F_{\mu\nu}F_{\rho\sigma}
+\frac{1}{2}k_{AF}^{\mu}\epsilon_{\mu\nu\rho\sigma}F^{\nu\rho}A^{\sigma}
-j^{\mu}A_{\mu}.
\end{equation}
The $k_{F}$ term is even under CPT, while the $k_{AF}$ term
(the Lorentz-violating Chern-Simons term of interst here) is odd under
CPT. The particular case of interest here is that of a timelike
$k_{AF}$. In fact, we shall consider a strictly timelike vector
$k_{AF}^{\mu}=(k,\vec{0}\,)$. The difficulty with this
theory is that the photon dispersion relation becomes
$\omega_{\pm}^{2}=p(p\mp 2k)$, where the $\pm$ denotes the helicities
of the modes.
At very long wavelengths, $p<|2k|$, one of the two helicity modes possesses an
imaginary frequency. As a result, there may be runaway solutions that grow
exponentially with time. In~\cite{ref-carroll1}, a Green's function was given that
alleviates this difficulty; the function provides solutions to the equations of
motion that do not involve exponential growth, but the cost is that charged particles
may begin to radiate before they are actually in motion.

The problem with runaway modes is also tied to the structure of the
energy-momentum tensor in the $k_{AF}$ theory. With only
the Chern-Simons term present, the purely electromagnetic part of the tensor
is~\cite{ref-carroll1}
\begin{equation}
\Theta^{\mu\nu}=-F^{\mu\alpha}F^{\nu}\,_{\alpha}+\frac{1}{4}g^{\mu\nu}
F^{\alpha\beta}F_{\alpha\beta}-\frac{1}{2}k_{AF}^{\nu}\epsilon^{\mu\alpha\beta\gamma}
F_{\beta\gamma}A_{\alpha}.
\end{equation}
With the strictly timelike $k_{AF}$, the energy density
($\Theta^{00}$), momentum density ($\Theta^{0j}$), and energy flux ($\Theta^{j0}$)
respectively are
\begin{eqnarray}
{\cal E} & = & \frac{1}{2}\vec{E}^{2}+\frac{1}{2}\vec{B}^{2}-
k\vec{B}\cdot\vec{A} \\
\vec{{\cal P}} & = & \vec{E}\times\vec{B} \\
\vec{S} & = & \vec{E}\times\vec{B}-kA_{0}\vec{B}+k\vec{A}\times\vec{E}.
\end{eqnarray}
These quantities are not gauge invariant,
although the integrated electromagnetic energy and momentum are.
Note, moreover, that ${\cal E}$ contains a term that is not bounded below;
this accounts for the existence of runaway solutions.

The question we want to address is whether a charge moving with
nonrelativistic speed $v$ in this vacuum will emit radiation. It
has already been established that there will be Cerenkov
radiation at higher order in $v$. For a charge with a finite speed, there will always be
a range of wavelengths for which the electromagnetic phase speeds are real and slower
than $v$, so Cerenkov radiation will be emitted at these wavelengths.

The algorithm
we previously developed for dealing with radiation into these modes~\cite{ref-altschul12}
gives the following results. The range of photon momenta $p$ satisfying
$0<\omega/p<v$ is $2k<p<2\gamma^{2}k$, where $\gamma=(1-v^{2})^{-1/2}$ is the usual
Lorentz factor. The (basically phase space) estimate for the radiated power
per unit wave vector is $P(\omega)=\frac{q^{2}}{8\pi}(1-\omega^{2}/p^{2}v^{2})\omega$.
So the total power emitted into the positive-frequency modes is
\begin{equation}
P=\frac{e^{2}}{8\pi}\int_{2k}^{2\gamma^{2}k}dp\left(1-\frac{1-2k/p}{v^{2}}\right)
p\sqrt{1-2k/p}.
\end{equation}
Changing variables to $u=(1-2k/p)/v^{2}$ (the square of the ratio of the phase speed to
the particle speed), this becomes,
\begin{eqnarray}
P & = & \frac{k^{2}e^{2}v^{3}}{2\pi}\int_{0}^{1}du\frac{\sqrt{u}(1-u)}{(1-v^{2}u)^{3}} \\
\label{eq-v3rad}
& = & \frac{k^{2}e^{2}v^{3}}{8\pi\gamma^{2}}\frac{v(3-v^{2})-(3+v^{2})(1-v^{2})\tanh^{-1}(v)}
{v^{5}} \\
& = & \frac{k^{2}e^{2}v^{3}}{30\pi}+{\cal O}(v^{5}).
\end{eqnarray}
[In spite of the $v^{5}$ in the denominator, the final fraction in (\ref{eq-v3rad}) is
regular at $v=0$.] The origin of the
overall $v^{3}$ dependence is fairly simple to understand;
the size of the $p$-space region covered by the integration is ${\cal O}(v^{2}k)$, and
the typical energy of photon emitted in this region is ${\cal O}(vk)$.

This shows the character of the radiation related to the modes with real frequencies.
However, this algorithm tells us nothing about what happens to the very longest
wavelength modes---those corresponding to the runaway solutions. It is entirely
plausible that these modes will be excited by a charged particle moving with even an
infinitesimal speed.

Therefore, we shall consider the possibility of radiation from a charge moving
with speed $v$ and only look at results up to ${\cal O}(v^{2})$. The purpose of this
restriction is
to disentangle the known Cerenkov radiation at shorter wavelengths from what happens
in the $p<|2k|$ modes. Fortunately, once this approximation is made, the ${\cal O}(v^{2})$
results may be determined using quite elementary methods. We shall also restrict
attention to results that are ${\cal O}(k^{2})$, which is the lowest order at which
nontrivial radiation effects could occur; in fact, for dimensional reasons alone,
the total energy loss rate must be proportional to $k^{2}$.
At these orders, we shall be able to show,
without needing to ``correct'' the theory with an acausal Green's function,
that a charged particle in uniform nonrelativistic motion does not emit any energy
as radiation.

The key to demonstrating this fact is calculating the ${\cal O}(k)$ contribution to
the magnetic field of the moving charge. For a charge $q$ located at the origin and
moving uniformly with a nonrelativistic velocity $\vec{v}=v\hat{z}$, the conventional
magnetic field is
\begin{equation}
\label{eq-B0}
\vec{B}_{0}(\vec{r}\,)=\frac{qv}{4\pi}\frac{\sin\theta}{r^{2}}\hat{\phi}.
\end{equation}
This is not the entirety of the magnetic field, however.
The standard $\vec{B}_{0}$ generates a further magnetic field through the
$k_{AF}$-modified Ampere's
Law (which is the only one of Maxwell's equations that is modified by the presence of
the purely timelike $k_{AF}$),
\begin{equation}
\vec{\nabla}\times\vec{B}-\frac{\partial\vec{E}}{\partial t}=2k\vec{B}+\vec{J}.
\end{equation}
We see that
the magnetic field itself behaves like an effective current source of strength
$\vec{J}_{{\rm eff}}=2k\vec{B}$.
It turns out that when $\vec{B}$ is given by (\ref{eq-B0}), the field of
the corresponding $\vec{J}_{{\rm eff}}$ may be found exactly.
This is obviously equivalent to the problem of finding the vector
potential of a charge in uniform motion, but in the Coulomb gauge
(since $\vec{\nabla}\cdot\vec{B}=0$), rather than the more usual Lorenz gauge.

An elementary version of the calculation proceeds as follows.
Between radii $r$ and $r+dr$ there is an effective
surface current $\vec{K}_{{\rm eff}}=\vec{J}_{{\rm eff}}\,dr$,
and the field of a surface current
$\vec{K}=K_{0}\sin\theta\,\hat{\phi}$ at radius $R$ is well known. (This $\vec{K}$
is, for example, the bound surface current on a uniformly magnetized sphere of radius $R$.)
The field generated by just this surface current is
\begin{equation}
\vec{B}_{K}(\vec{r}\,)=\left\{
\begin{array}{ll}
\frac{2}{3}K_{0}\hat{z}, & r<R \\
\frac{R^{3}}{3r^{3}}K_{0}(3\cos\theta\hat{r}-\hat{z}), & r>R \\
\end{array}
\right.
\end{equation}

These contributions to the total magnetic field
field must be integrated over all the shells from radii $R=0$ to $\infty$,
remembering that the equatorial surface current density $K_{0}$ is also
a function of $R$.
The required integral is
\begin{eqnarray}
\vec{B}(\vec{r}\,) & = & \vec{B}_{0}(\vec{r}\,)+\int_{0}^{r}dR\left(\frac{kqv}
{2\pi R^{2}}\right)\left[\frac{R^{3}}{3r^{3}}(3\cos\theta\,\hat{r}-\hat{z})
\right]+\int_{r}^{\infty}dR\left(\frac{kqv}{2\pi R^{2}}\right)\left(\frac{2}{3}\hat{z}
\right) \\
& = & \vec{B}_{0}(\vec{r}\,)+\frac{kqv}{4\pi r}(\cos\theta\,\hat{r}+\hat{z}) \\
\label{eq-B1}
& = & \vec{B}_{0}(\vec{r}\,)+\frac{kqv}{4\pi r}(2\cos\theta\,\hat{r}-
\sin\theta\,\hat{\theta}).
\end{eqnarray}
Because of the presence of the dimensional parameter $k$, this field decays only
as $r^{-1}$---less rapidly than the conventional field $\vec{B}_{0}$. This means
that the field at ${\cal O}(k)$ is potentially capable of producing outgoing energy
and momentum fluxes at infinity.

To see whether there really are such fluxes, we must turn our attention to the
energetics of the fields at large $r$. We shall look specifically at the energy
flux $\vec{S}$, which is a modified Poynting vector. An outward energy flux at large
$r$ will be the signature of energy losses from vacuum Cerenkov radiation.

We recall that we are only interested in effects at ${\cal O}(v^{2})$, since
the presence of Cerenkov radiation into real-frequency modes at ${\cal O}(v^{3})$
has already been established. There are three terms in $\vec{S}$, and two of them
involve cross products with $\vec{E}$. However, neither of these can contribute
to a flux at infinity at this order. The standard Coulomb field of the charge points
radially outward, and so a cross product containing the conventional part of $\vec{E}$
can never contribute to an outgoing flux. The magnetic quantities
$\vec{A}$ and $\vec{B}$ are each proportional to $v$, and the inductive part
of $\vec{E}$ [the part sourced by $\partial B/\partial t=v(\partial B/\partial z)$]
is ${\cal O}(v^{2})$; so the products $(\vec{E}\times\vec{B})\cdot\hat{r}$ and
$(\vec{A}\times\vec{E})\cdot\hat{r}$ are necessarily ${\cal O}(v^{3})$.
Then the only term in $\vec{S}$ that remains a concern is $-kA_{0}\vec{B}$.
Since this term contains an explicit factor of $k$, we
evidently do not need to consider $\vec{B}$ beyond ${\cal O}(k)$ in order to
determine the ${\cal O}(k^{2})$ flux. Any $k$-dependent contributions to $A_{0}$
would also be ${\cal O}(v^{2})$ (as are the next $k$-independent contributions to
$A_{0}$), and so these may be neglected, leaving
the only contribution as that coming from the $k$-dependent $\vec{B}$ (\ref{eq-B1}).

The radial flux at spatial infinity is thus
\begin{equation}
\vec{S}\cdot\hat{r}=-k\left(\frac{q}{4\pi r}\right)
\left[\frac{kqv}{4\pi r}(2\cos\theta\,\hat{r}-\sin\theta\,\hat{\theta})\right]
\cdot\hat{r}=-\frac{k^{2}q^{2}v}{8\pi^{2}r^{2}}\cos\theta.
\end{equation}
This is not, on its own, a meaningful result. There
naively appears to be a constant energy flux
along the $-z$-direction. However, we must recall that local expressions for energy
densities are not gauge invariant in this theory. What is gauge invariant, however, is
the integral of this flux over the full surface $\Sigma$ at spatial infinity, and
\begin{equation}
\int_{r\rightarrow\infty}d\Sigma\,(\vec{S}\cdot\hat{r})=0.
\end{equation}
There is no net energy flux away from the particle---no vacuum Cerenkov radiation.

This can be immediately extended to show that there is no momentum flux at infinity
at ${\cal O}(v)$
either. The momentum flux involves $\Theta^{jk}$, which takes its standard form
when $k_{AF}$ is purely timelike. Because the Maxwell stress terms only
involve products $E_{j}E_{k}$ and $B_{j}B_{k}$, there are no terms that are
linear in
$v$ that do not fall off faster than $r^{-2}$. So there are no energy or momentum
fluxes at spatial infinity at the lowest orders in $v$ and $k$.

If $k_{AF}$ is not purely timelike, but $|\vec{k}_{AF}|$ is small compared with
$k_{AF}^{0}$, then it is possible to boost into a frame in which $k_{AF}'$ is
indeed purely timelike, without giving the charged particle a relativistic
velocity in that frame. The speed added to the particle is
$v'=|\vec{k}_{AF}|/k_{AF}^{0}$; the absence of vacuum Cerenkov radiation for a
particle moving with this speed in the boosted
frame---a result which is valid to
${\cal O}[(k_{AF}^{0})^{2}v']$---ensures that there is no radiation from a
stationary particle to ${\cal O}(|\vec{k}_{AF}|k_{AF}^{0})$. It follows from
the series expansion that
there is no contribution to the emission rate at this particular order, even if
$|\vec{k}_{AF}|$ and $k_{AF}^{0}$ are comparable in size. This mirrors results
from~\cite{ref-lehnert2}, in which it was found that there was no radiation
from a stationary particle at this order in the
alternative $|\vec{k}_{AF}|>|k_{AF}^{0}|$
spacelike case.

We should note, however,
that it is by no means a triviality that a stationary particle should emit no
vacuum Cerenkov radiation. In fact, in the spacelike case, such radiation does occur,
albeit not at the low order just discussed. When the $k_{AF}$ is anisotropic, it
provides a preferred spatial direction along which energy and momentum may be
emitted.

It is similarly non-obvious that a charge moving in an isotropic $k_{AF}$
background should not emit radiation at ${\cal O}(k^{2}v^{2})$. There is no obvious
symmetry that could prevent the emission (except Lorentz symmetry, which has
been explicitly broken). The $k$ term is odd under P and CPT, but this does
not preclude a P-even radiation process from occurring at ${\cal O}(k^{2})$.
The situation is not isotropic, because the axis
of motion $\vec{v}$ picks out a particular direction, and it would not
be surprising if the charge lost energy and momentum to electromagnetic waves
emitted into a cone of angles around $\hat{v}$. This is precisely what happens
at higher order in $v$, through the interaction with non-runaway modes.

The main result of this paper has been the demonstration that a charged
particle moving slowly through a vacuum that violates Lorentz boost and CPT
symmetries (but not spatial isotropy) does not emit any radiation at ${\cal O}(v^{2})$.
While there is radiation at ${\cal O}(v^{3})$, it can
be understood as occurring primarily because energy-momentum conservation allows it.
When the phase speed for a given mode of the electromagnetic field is less than
the speed at which a charged particle is moving, Cerenkov radiation is expected,
in analogy with the emission by relativistic particles passing through material with
a significant index of refraction. However, things are qualitatively different for
the long-wavelength modes of the isotropic theory. These modes lead to the possibility
of runaway solutions, and their contributions to vacuum Cerenkov radiation could
not have
been evaluated by any of the methods that had previously been introduced to study
this kind of radiation process in other regimes.

So this paper has dealt with the
greatest remaining puzzle that had arisen in studies of
Lorentz-violating vacuum Cerenkov radiation. The central
result is simple and aesthetically interesting on several levels. For instance, 
as a side effect of this calculation, we determined the magnetic field of a
nonrelativistically moving charge in the Lorentz-violating theory, to leading order
in $k$. This was itself an interesting result, since it turned out the field
could be built up from the interior and exterior fields
of a set of concentric shells.

On a deeper level, this work also addresses the direct problem of the instability
in the timelike $k_{AF}$ theory. As we noted, runaway modes can be wiped out of
the theory by using a Green's function for the electromagnetic field that is
acausal---the radiation at a time $t$ depending on the behavior of the source
at times $t'>t$. This cures one problem by introducing another. However, we saw in
this paper that the runaway modes were not a problem in the context being studied.
We showed that no acausal signaling was required to avoid exponentially growing
energy-carrying solutions. Although this was obviously restricted to a very
particular regime, it provides an important insight into the behavior of this
fascinating Lorentz-violating theory.

\end{document}